\documentclass[aps,pre,twocolumn,showpacs,superscriptaddress,groupedaddress]{revtex4-1}
\UseRawInputEncoding
\pdfoutput=1
\usepackage{hyperref}
\usepackage{epsfig}
\usepackage{amsmath}
\usepackage{graphicx}
\usepackage{wrapfig}
\usepackage{dcolumn}
\usepackage{bm}
\usepackage{amssymb}
\usepackage{titlesec}
\usepackage{mathtools}
\usepackage{relsize}
\usepackage{cleveref}
\usepackage{calligra}
\usepackage[usenames,dvipsnames]{color}

\def\corr{\textcolor{black}}

\begin{document}
\title{Critical transition for colliding swarms}
\author{Jason Hindes$^{1}$, Victoria Edwards$^{1,2}$,M. Ani Hsieh$^2$ and Ira B. Schwartz$^{1}$}
\affiliation{$^{1}$U.S. Naval Research Laboratory, Washington, DC 20375, USA\\
$^2$Mechanical Engineering and Applied Mechanics, University of Pennsylvania, Philadelphia, Pennsylvania 19104}

\begin{abstract}
Swarming patterns that emerge from the interaction of many mobile agents are a subject of great interest in fields ranging from biology to physics and robotics. 
In some application areas, multiple swarms effectively interact and collide, producing complex spatiotemporal patterns. Recent studies have begun to address swarm-on-swarm dynamics, and in particular the scattering of two large, colliding swarms with nonlinear interactions. To build on early numerical insights, we develop a self-propelled, rigid-body approximation that can be used to predict the parameters under which colliding swarms are expected to form a milling state. Our analytical method relies on the assumption that, upon collision, two swarms oscillate near a limit-cycle, where each swarm rotates around the other while maintaining an approximately constant and uniform density. Using this approach we are able to predict the critical swarm-on-swarm interaction coupling, below which two colliding swarms merely scatter, as a function of physical swarm parameters. \corr{We show that the critical coupling gives a lower-bound for all impact parameters, including head-on collision,} and corresponds to a saddle-node bifurcation of a stable limit cycle in the uniform, constant density approximation. Our results are tested and found to agree with both small and large multi-agent simulations.
\end{abstract}
\maketitle

\section{\label{sec:Intro}INTRODUCTION}
Swarming occurs when spatiotemporal patterns and behaviors emerge from the interaction of large numbers of coupled mobile systems, typically 
with fairly limited capabilities and local dynamics. Examples have been
discovered in nature over many spatiotemporal scales from colonies of bacteria, to swarms of
insects\cite{Theraulaz2002,Topaz2012,Polezhaev,Li_Sayed_2012}, flocks of birds \cite{Leonard2013,Ballerini08,Cavagna2015}, 
schools of fish\cite{Couzin2013, Calovi2014},
crowds of people\cite{Rio_Warren_2014}, and active-matter systems more generally\cite{Cichos2020}.
Understanding the principles behind swarming patterns and describing how they emerge from simple models has been the subject of significant work in physics, applied mathematics, and engineering sciences \cite{Vicsek,Marchetti,Aldana,Solon2015,PhysRevLett.117.038103,PhysRevLett.121.178001,PhysRevLett.122.258001,Desai01, Jadbabaie03, Tanner03b, Tanner03a,  Gazi05, Tanner07}. 
Parallel with this work, and because of the robustness, scalability, and collective-problem solving capabilities of natural swarms, much research has focused on 
designing and building swarms of mobile robots with a large and ever expanding number of platforms, as well as virtual and physical interaction mechanisms\cite{Cichos2020,AutonomousMobileRobots,8990018,7989200,MultiRobotSystems,4209425}. 
Applications for such systems range from exploration\cite{8990018},
mapping\cite{Ramachandran2018}, resource allocation \cite{Li17, Berman07,
  Hsieh2008}, and swarms for defense \cite{Wong2020,Chung2011, Witkowski}

Since the overall cost of robotic systems has decreased significantly in recent years,
it has become possible to use artificial swarms in the real world \cite{6224638,TERMES,7989200,8990018}.
This introduces the possibility of having multiple swarms occupying the 
same physical space, resulting in mutual interactions and perturbations of one another's 
dynamics\cite{Sartoretti}. As the potential for such swarm-on-swarm interactions increases, a
basic physical understanding of how multiple swarms collide and merge becomes necessary. 
\corr{Recent work in swarm robotics and autonomy has begun to address how swarms  
can be designed to detect, herd, or capture another\cite{8444217,9029573,9303837}. Yet, 
most approaches are algorithmic and rely on simulation-based optimization, and 
are thus lacking in basic physical and analytical insights.} 


Though much is known about the behaviors and stability of
single isolated swarms with physically-inspired, nonlinear interactions\cite{Levine,Erdmann,Minguzzi,DOrsagna,Romero2012},
much less is known about the intersecting dynamics of multiple such
swarms, even in the case where one swarm is a single particle, as in predator--prey modeling\cite{Chen2014}. 
Recent numerical studies have shown that when two flocking swarms collide,
the resulting dynamics typically appear as a merging of the swarms into a single flock, 
milling as one uniform swarm, or scattering into separate composite flocks moving in different directions\cite{ARMBRUSTER201745,kolon2018dynamics,Sartoretti}. 
Though interesting, a more detailed analytical understanding of how and when these behaviors occur is needed, \corr{especially when designing 
robotic swarm experiments for, e.g., swarm herding and capture\cite{9029573,9303837,8444217}, and controlling their outcomes.}

To make progress, we consider \corr{a very well-known model of swarming\cite{Levine,DOrsagna,Erdmann,Minguzzi,Bernoff,AlbiStability2014}}, consisting of mobile agents moving under the influence of self-propulsion, nonlinear damping, and pairwise interaction forces. In the absence of interactions, each swarmer tends to a fixed speed, which balances its self-propulsion and damping, but has no preferred direction\cite{hindes2020stability}. A simple model 
that captures the basic physics is
\begin{align}
\ddot{\bold{r}}_{i}= \big[\alpha_{i} -\beta|\dot{\bold{r}}_{i}|^{2}\big]\dot{\bold{r}}_{i}-\lambda_{i}\sum_{j\neq i} \partial_{\bold{r}_{i}}U(|\bold{r}_{j}-\bold{r}_{i}|)
\label{eq:SwarmModel}
\end{align} 
where $\bold{r}_{i}$ is the position-vector for the $i$th agent in two spatial dimensions, $\alpha_{i}$ is a self-propulsion constant, $\beta$ is a damping constant, and $\lambda_{i}$ is a coupling constant\cite{Levine,Erdmann,Minguzzi,DOrsagna,Romero2012}. The total number of swarming agents is $N$, and each agent has unit mass. Beyond providing a basis for theoretical insights, Eq.(\ref{eq:SwarmModel}) has been implemented in experiments with several robotics platforms including autonomous ground, surface, and aerial vehicles\cite{s4209425,zwaykowska2016collective,edwards2020delay,hindes2020unstable}.

An example interaction potential that we consider in detail is the Morse potential, 
\begin{align}
U(r)=Ce^{-r/l}-e^{-r}
\label{eq:Morse}
\end{align} 
-- a common model for soft-core interactions with local repulsion and
attraction ranges, scaled as $l$ and $1$, respectively\cite{DOrsagna,Bernoff}. 
In the following, we assume that two interacting swarms are subject to the same underlying physics,
Eqs.(\ref{eq:SwarmModel}-\ref{eq:Morse}), but with different initial conditions and potentially different control parameters. In particular, we assume that 
within each swarm the parameters are homogeneous, e.g., $\alpha_{i}\in\{\alpha^{(1)},\alpha^{(2)}\}$ and $\lambda_{i}\in\{\lambda^{(1)},\lambda^{(2)}\}$, where the superscripts (1) and (2) denote the first and second swarms, respectively. \corr{The summation in Eq.(\ref{eq:SwarmModel}) is taken over all agents (in both swarms). However, by construction, each swarm will be initially separated by a large distance compared to the interaction scales, $l$ and $1$, and the maximum distance between agents within each swarm. Therefore, the interaction force an agent feels will be at early times effectively confined to their own swarm, given the exponential decay with distance implied by Eq.(\ref{eq:Morse}).} \corr{The assumption that the two swarms satisfy the same basic physics makes sense if the swarms are composed of similar agents, and should be a reasonable, baseline assumption for the collision of swarms of simple, programmable mobile robots.}  

\section{\label{sec:Collision}COLLISION OF TWO FLOCKING SWARMS}
As in \cite{ARMBRUSTER201745,kolon2018dynamics}, we are interested in the
collision of two flocking swarms composed of approximately equal numbers of
agents. The swarms are each prepared at $t\!=\!0$ in a flocking state with initial velocities and positions that are a large distance from the collision region \corr{($D\!=\!50\!\gg\!1,\;l,\;\text{and the sizes of the flocks}$)}, such that $\bold{r}_{i}\!=\!\bold{d}_{i}^{(1)}\!-\!D\;\bold{\hat{x}}$ and $\dot{\bold{r}}_{i}=\sqrt{\alpha^{(1)}/\beta}\;\bold{\hat{x}}$ if $i\in(1)$, and 
\corr{$\bold{r}_{i}\!=\!\bold{d}_{i}^{(2)}\!+\!D\sqrt{\alpha^{(2)}/\alpha^{(1)}}\bold{\hat{x}}+\Delta y\;\bold{\hat{y}}$ and $\dot{\bold{r}}_{i}=-\sqrt{\alpha^{(2)}/\beta}\;\bold{\hat{x}}$ if $i\in(2)$. The internal flocking coordinates, $\bold{d}_{i}^{(1)}$ and $\bold{d}_{i}^{(2)}$}, represent local minimum energy configurations \corr{(LMECs)}, defined by \corr{$-\sum_{j\neq i}\partial_{\bold{d}_{i}}U(|\bold{d}_{j}-\bold{d}_{i}|)=\bold{0}_{i}$} $\forall i$ \cite{CARRILLO2014332}. \corr{This property demonstrates one of the advantages of models like Eq.(\ref{eq:SwarmModel}) for running robotic-swarm experiments, since the relative configurations of flocks can be directly controlled through the potential function.} \corr{Note that the speed of the $i$th agent is equal to $\sqrt{\alpha_{i}/\beta}$, which is the condition that allows for flocking.} Given this setup, the net force on every agent is initially zero-- a consequence of the LMEC and the finite-range of interactions. 

\corr{In general, in two spatial dimensions there are four initial conditions that one can specify for the centers of each flock. However, note that the speeds are fixed by the flocking condition, and the absolute positions do not matter, only the relative distances-- leaving two initial-condition parameters, the relative distance and velocity between the flocks. In this work, however, we are interested in collisions that result in swarm milling states, which have zero total linear momentum. As a consequence, the flocks should be nearly aligned upon collision, with small transverse velocities. In the nearly aligned regime, the relevant initial-condition parameter is the distance between the two flocks as they approach $x\!=\!0$, regardless of the direction of their velocities. This distance is often called the ``impact parameter" in classical mechanics\cite{Landau1976Mechanics}, and it signifies the closest distance the two flocks would approach in the absence of interaction forces. The impact parameter is denoted $\Delta y$ in our initial conditions.} 

\begin{figure}[h]
\center{\includegraphics[scale=0.22]{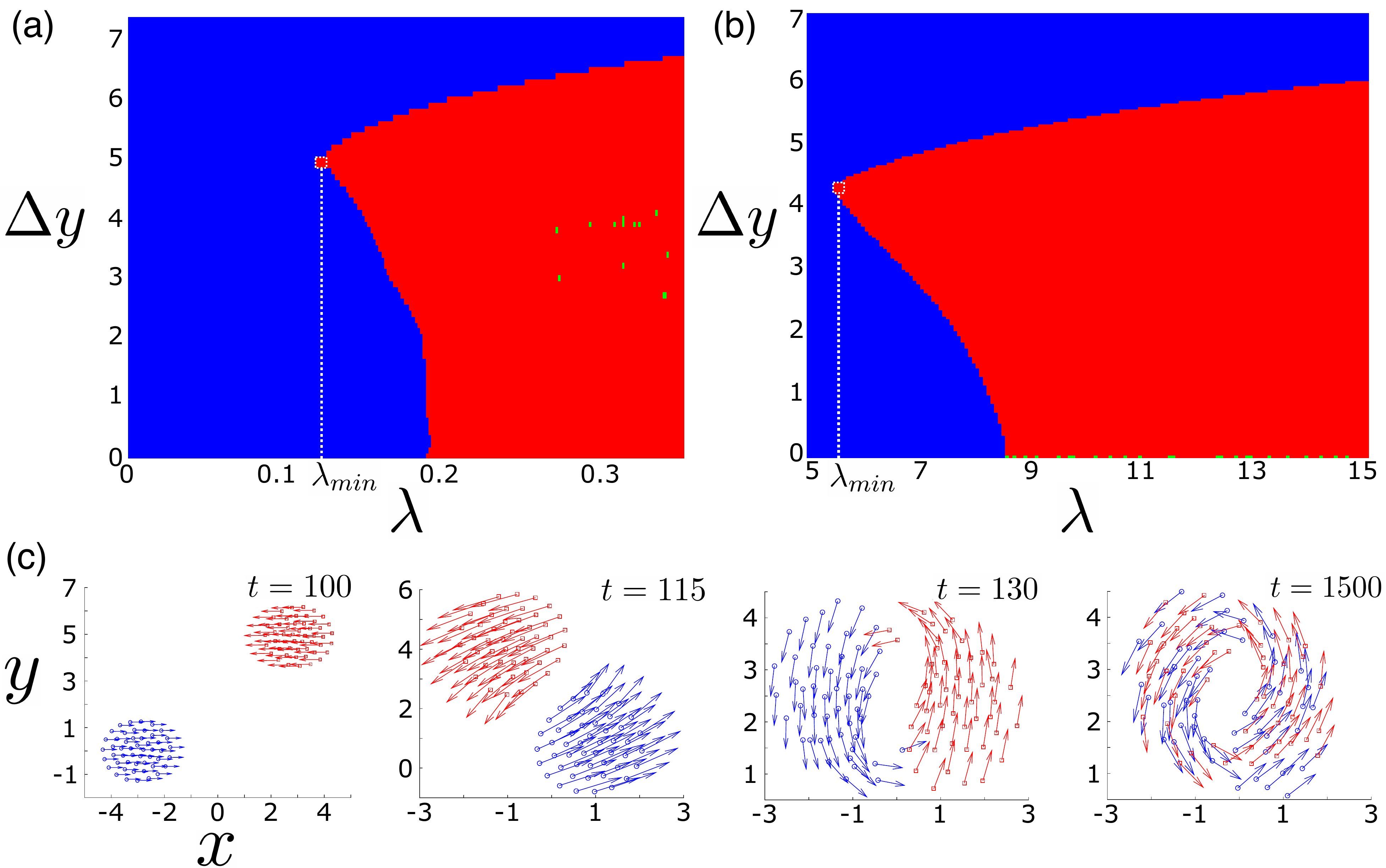}}
\caption{Collision of two symmetric flocks. (a) Scattering digram indicating the aggregate swarm state \corr{as a function of the distance between the two flocks before collision $\Delta y$, and the coupling $\lambda$}: scattering (blue), milling (red), and merged flocking (green). The critical coupling is specified with a dashed vertical line, and separates the scattering and milling regions. \corr{(b) The same diagram for the collision of two agents.} (c) Four time-snapshots for $\lambda\!=\!\lambda_{min}$ showing each swarm with different colors: red squares and blue circles. Velocities are drawn with arrows. Swarm parameters are $\alpha\!=\!1$, $\beta\!=\!5$, $C\!=\!10/9$, $l\!=\!0.75$, and $N\!=\!100$. \corr{The swarms are initially separated by a large distance, $2D\!=\!100$ for all simulations.}}
\label{fig:Shedding1}
\end{figure}

\corr{Depending on the value of $\Delta y$} and the coupling strength, the two flocks typically scatter or mill. In the former the swarms leave the collision region in separate flocking states with perturbed velocities. In the latter they form a milling state (MS), and circulate around a stationary center of mass\cite{Costanzo_2018}. \corr{To guide our analysis, we perform numerical experiments for different values of $\Delta y$ and $\lambda$, and determine which final state the swarms relax to}. Figure \ref{fig:Shedding1}(a) shows such an example, final-scattering diagram for the collision of symmetric flocks with equal parameters. The final swarm states are specified with blue and red for scattering and milling, respectively; the green portions indicate the formation of a combined flocking state, which is comparatively infrequent for the parameters shown (and decreases in frequency as $N\!\rightarrow\!\infty$). \corr{The scattering diagram is built by integrating Eq.(\ref{eq:SwarmModel}) from the initial separation until well after collision, $t\!=\!1500$. For a fixed value of $\Delta y$, $\lambda$ is swept from $0.01$ to $0.35$ in increments of $0.0017$. Similarly, $\Delta y$ is swept from $0$ to $7.5$ in steps of $0.1$. For each value of $\Delta y$ and $\lambda$, if the center of mass (CM) of both swarms is stationary, the point is colored red. If the two CMs continue to separate after collision, the point is colored blue. If the CMs are not stationary but remain a fixed distance apart, the point is colored green.}

In general, predicting the complete scattering diagrams for such swarm collisions is a very hard nonlinear dynamics problem, since ultimately one must address whether or not a certain set of initial conditions in a high-dimensional phase space falls within the basin of attraction for a given final state. \corr{This kind of question does not typically have a systematic answer, particularly for high-dimensional nonlinear systems, apart from simply doing many numerical simulations.} \corr{For instance, if the collisions are head-on ($\Delta y\!=\!0$), then $\lambda\!\gtrsim\!0.19$ in order for a MS to form. On the other hand, if $\Delta y$ is much larger than the attractive length scale ($1$), then the coupling must be similarly large; e.g., if $\Delta y\!=\!7$, then $\lambda\!\gtrsim\!0.35$ (keep in mind that swarms can very much collide even if $\Delta y$ is larger than the nominal sizes of the flocks, since Fig.\ref{fig:Shedding1} (a) shows red regions for sufficiently large couplings). Hence, whether or not a MS is formed clearly depends on initial conditions-- different $\Delta y$ have different, non-generic, transition couplings. However, between the two limits we find a unique critical value $\lambda_{min}$, designated with a white box and dashed line in Fig.\ref{fig:Shedding1}(a), which is the smallest coupling needed to form a MS, over all impact parameters, $\Delta y$. For reference, this value is $\lambda_{min}\!=\!0.13$, or approximately $0.7$-times the head-on-collision value in Fig. \ref{fig:Shedding1}(a).} 

\corr{What the critical point might mean, in general, can be understood by first studying the scattering-diagram for a simple two-agent system, shown in Fig.\ref{fig:Shedding1} (b). Note that the two diagrams are qualitatively very similar despite the significant difference in the number of agents. Even in the simple two-agent case, it is not clear that the scattering-diagram can be derived without resorting to simulations. However, it is possible to show that the MS corresponds to a stable limit-cycle oscillation that is born in a generic saddle-node (SN) bifurcation, exactly at the point where the red-region first emerges as we sweep the coupling in Fig.\ref{fig:Shedding1} (b) (see App.\ref{sec:App1}). Crucially, the SN bifurcation is independent of the impact-parameters $\Delta y$, and in fact, gives a rigorous lower bound for the transition coupling for all impact parameters in the system. Building on this bifurcation insight we focus on $\lambda_{min}$ in this work, because it is generic, in that it depends on the physical parameters of the swarm and not on initial conditions, and it gives a lower-bound for head-on collisions and all other collision distances $\Delta y$. Just as in the two-agent case, we will show that $\lambda_{min}$ corresponds to the birth of stable bound-state oscillations of two flocks, and therefore we can do a first-principles calculation to approximate it.}

In order to visualize collisions that result in milling in many-agent swarms, we show four
time-snapshots in Figure \ref{fig:Shedding1}(c) when
$\lambda\!=\!\lambda_{min}$. Agents in the two swarms are drawn with different
colors, and their velocities shown with arrows. In the first snapshot (upper
left), the swarms approach collision with configurations and velocities
identical to those specified in the first paragraph of this section-- namely,
the LMEC with constant velocity. In the second snapshot (upper right) the
swarms rotate around each other with a constantly changing heading, roughly
uniform velocity distribution, and a configuration approximately equal to the
LMEC. Over time each swarm's density elongates in the direction of rotation
(third snapshot, lower left), as the velocity distribution becomes less
homogeneous. Finally, on long times scales the two swarms blend into one and
form a MS with agents from each uniformly distributed across the whole.

In order to predict the critical coupling, $\lambda_{min}$, our approach is to find an analytical description of the collision dynamics that is applicable for the first two snapshots in Figure \ref{fig:Shedding1}(c), where two approximately \corr{constant-density} flocks approach, and then rotate around a common center. Our conjecture is that if such rotations are approximately stable, then a MS occurs upon collision (and visa versa). Though we will analyze two-flock collisions in this way assuming Eq.(\ref{eq:Morse}), our method should be applicable to a broader range of second-order dynamical swarms given position-dependent, nonlinear interactions with finite attractive and repulsive length scales. 

\subsection{\label{sec:UCDA} Uniform constant density approximation}
\begin{figure}[h]
\center{\includegraphics[scale=0.219]{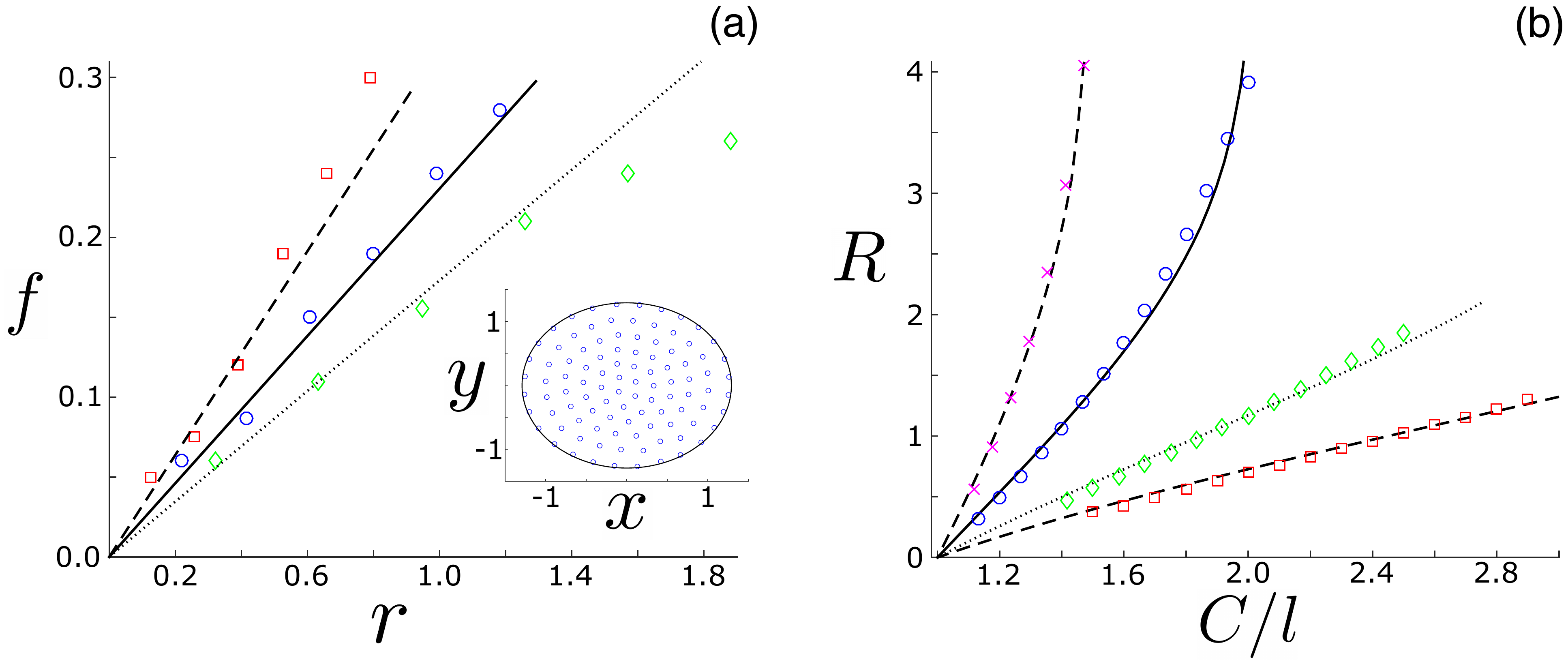}}
\caption{Uniform constant density approximation for flocking states (UCDA). (a) fraction of agents a distance $r$ from the flock's center for $C\!=\! 1.0\; (\text{red-squares}),\; 1.1\; (\text{blue-circles}),\; 1.25\; (\text{green-diamonds})$ when $l\!=\!0.75$, where $C$ is the repulsive-force strength with length-scale $l$. The dashed, solid, and dotted lines indicate UCDA predictions from solving Eq.(\ref{eq:R}). The inlet panel shows an example flocking state with the UCDA boundary drawn in black for $C\!=\!1.1$. (b) Flocking state boundary, $R\!=\!\max\{r\}$, from simulations: ($l\!=\!0.85$, magenta-xs), ($l\!=\!0.75$, blue-circles), ($l\!=\!0.60$, green-diamonds) and ($l\!=\!0.50$, red-squares) compared to UCDA predictions shown with lines near each series. Other swarm parameters are $\alpha\!=\!1$, $\beta\!=\!5$, $\lambda\!=\!2$, and \corr{$N_{1}\!=\!100$}.} 
\label{fig:Shedding2}
\end{figure}

First, we would like to find a low-dimensional approximation for the flocking state dynamics. A clue comes from Figure \ref{fig:Shedding2}(a), which plots the fraction of agents at a given distance $r$ from the CM of a single moving flock for different values of the repulsion strength, $C$. We can see that the radial distribution is approximately {\it linear} in $r$. Moreover, since the potential is radial, we expect the steady-state angular distribution to be uniform; the inlet panel shows an example flocking state with such a spatial distribution of agents. Together, these imply a roughly uniform density in the flocking state, $\rho\!=\!N_{1}/\pi R^{2}$, where $R$ is the maximum radius \corr{and $N_{1}$ is the number of agents in flock (1)}. Given the uniform-density assumption, the predicted fraction of agents at a given $r$ is $f(r)\!=\!2r\Delta r/R^{2}$, where $\Delta r$ is the bin-size used to plot the distribution. This prediction is drawn with lines for comparison in Figure \ref{fig:Shedding2}(a). \corr{Note that the actual distribution is not-quite linear in $r$; the uniform density approximation predicts both more and fewer agents near the flock's boundary than is actually observed, depending on the value of $C$. This suggests a straightforward improvement in the accuracy of our calculations (that follow): input the {\it exact} density function.}

Assuming a uniform density, we can describe a flock in general by the position of its center, the velocity, and the boundary radius, $R$. In particular, every agent, including those on the boundary, move with constant speed, $\sqrt{\alpha/\beta}$, where $\alpha$ is the self-propulsion constant for the flock. A self-consistent formula can be derived for $R$, and used to compute it, by satisfying force-balance on the boundary. For example, consider an agent with $\bold{d}_{i}\!=\!R\;\hat{\bold{x}}$. The $x$-component of the interaction force must be zero,  
\begin{align}
0=\int_{0}^{2\pi}\!\!\!\int_{0}^{1}\!&\Bigg(\!\frac{C}{l}e^{-\frac{R}{l}\sqrt{1+u^{2}-2u\!\cos{\phi}}}-e^{-R\sqrt{1+u^{2}-2u\!\cos{\phi}}}\Bigg) \nonumber \\
&\cdot \frac{u\!\cos{\phi}-1}{\sqrt{1+u^{2}-2u\!\cos{\phi}}}\cdot u du d\phi\;, 
\label{eq:R}
\end{align} 
where $u\!\equiv\!r/R$. Note that the $y$-component of the force is trivially zero due to the uniform-angular distribution of agents. Comparisons between simulations and numerical solutions to Eq.(\ref{eq:R}) are shown in Fig.\ref{fig:Shedding2}(b) for a range of control parameters, and indicate good agreement. \corr{Because of this agreement, and its relative simplicity, we continue our analysis assuming a uniform steady-state density of agents in a flock.}

Next, we can approximate the initial collision dynamics of the flocks by assuming that the uniform density configuration remains constant within each flock, with a boundary given by Eq.(\ref{eq:R}). Namely, the rigid-body collision model that we will analyze below is of two interacting, constant-density disks composed of self-propelled agents. Consider two representative agents positioned at the center of each swarm, $\bold{r}^{(1)}(t)$ and $\bold{r}^{(2)}(t)$. \corr{If we directly apply Eq.(\ref{eq:SwarmModel}), it is easy to check that the interaction forces on agents positioned at $\bold{r}^{(1)}(t)$ and $\bold{r}^{(2)}(t)$, due only to agents within the same flock, vanish, given the assumed uniform angular distribution of agents within each flock. If agents away from the center of flock $(1)$ have coordinates $\bold{r}_{j}^{(1)}\!=\!\bold{r}^{(1)}\!+r\cos{\phi}\hat{\bold{x}}+r\sin{\phi}\hat{\bold{y}}$, then the interaction force on an agent at $\bold{r}^{(1)}$ from flock $(1)$ is: 
\begin{align}
\bold{0}=\int_{0}^{2\pi}\!\!\!\int_{0}^{R}\!&\bigg(\!\frac{C}{l}e^{-r/l}-e^{-r}\!\bigg)\frac{N_{1}rdrd\phi}{\pi R^{2}}\big(\!\cos{\phi}\hat{\bold{x}}+\sin{\phi}\hat{\bold{y}}\big). 
\end{align}}    
Hence, the non-zero contributions to the interaction sums in Eq.(\ref{eq:SwarmModel}) for $\bold{r}^{(1)}(t)$ and $\bold{r}^{(2)}(t)$ only come from the {\it other flock}, since the interaction force from their own flocks cancel. Moreover, the interaction force from the opposing flock is felt gradually as the two swarms approach, because of the finite-range interactions and the initially large separation between the flocks. To find the non-zero contribution, we simply need to integrate the interaction force over a constant-density disk of radius $R$, centered on the opposing swarm's center, $\bold{r}^{(2)}(t)$ or $\bold{r}^{(1)}(t)$, respectively. If we assume that the two swarms are equally sized, each with $N/2$ agents, \corr{directly applying Eq.(\ref{eq:SwarmModel}) for $\bold{r}^{(1)}(t)$ and $\bold{r}^{(2)}(t)$ gives:} 
\begin{subequations}
\begin{align}
\label{eq:UCDA1}
&\ddot{\bold{r}}^{(1)}= \big[\alpha^{(1)} \!-\beta|\dot{\bold{r}}^{(1)}|^{2}\;\big]\dot{\bold{r}}^{(1)} -\frac{\lambda^{(1)}N}{2}\boldsymbol{\mathcal{E}}(\bold{r}^{(2)}\!,\bold{r}^{(1)};R)\\
\label{eq:UCDA2} 
&\ddot{\bold{r}}^{(2)}= \big[\alpha^{(2)} \!-\beta|\dot{\bold{r}}^{(2)}|^{2}\;\big]\dot{\bold{r}}^{(2)} -\frac{\lambda^{(2)}N}{2}\boldsymbol{\mathcal{E}}(\bold{r}^{(1)},\bold{r}^{(2)};R)\\
\label{eq:UCDA3}
&\boldsymbol{\mathcal{E}}(\bold{r}^{(2)}\!,\bold{r}^{(1)};R)=\int_{0}^{2\pi}\!\!\!\int_{0}^{R}\!\! \frac{\bold{r}^{(2)}+\bold{d}-\bold{r}^{(1)}}{|\bold{r}^{(2)}+\bold{d}-\bold{r}^{(1)}|}\cdot \frac{r dr d\phi}{\pi R^{2}} \nonumber \\
&\;\;\;\;\;\;\;\;\;\;\;\;\;\;\cdot\Bigg(\frac{C}{l}e^{-|\bold{r}^{(2)}+\bold{d}-\bold{r}^{(1)}|/l}-e^{-|\bold{r}^{(2)}+\bold{d}-\bold{r}^{(1)}|}\Bigg)\\
&\bold{d}= r\cos{\phi}\;\hat{\bold{x}} \;+\; r\sin{\phi}\;\hat{\bold{y}},  
\label{eq:UCDA4}
\end{align}
\end{subequations}
where $\bold{d}$ is an internal-coordinate inside the constant-density disk centered at $\bold{r}^{(2)}(t)$ or $\bold{r}^{(1)}(t)$, respectively. Equations (\ref{eq:R}-\ref{eq:UCDA4}) constitute the rigid-body dynamical system that we call the uniform constant density approximation (UCDA) \corr{\cite{NoteColl2}}. The integrals in Eq.(\ref{eq:UCDA3}) can be evaluated using e.g., trapezoid rule. Our next step is to study stable oscillations of ${\bold{r}}^{(1)}(t)$ and ${\bold{r}}^{(2)}(t)$ in the UCDA and compare to swarm collision dynamics.

\subsection{\label{sec:Oscillations} Stable Oscillations}
Stable oscillations in the UCDA come in the form of circular-orbit limit cycles where both flocks oscillate around a common center with the same frequency, a fixed phase difference, and different 
amplitudes in general. We can compute the parameters for such limit cycles by
substituting the ansatz $\bold{r}^{(1)}(t)\!=\!A_{1}\!\cos(\omega
t)\hat{\bold{x}}+A_{1}\!\sin(\omega t)\hat{\bold{y}}$ and
$\bold{r}^{(2)}(t)\!=\!A_{2}\!\cos(\omega
t+\gamma)\hat{\bold{x}}+A_{2}\!\sin(\omega t+\gamma)\hat{\bold{y}}$ into
Eqs.(\ref{eq:UCDA1}-\ref{eq:UCDA4}). The result is the following four
root equations satisfying $F_i\!=\!0$ for $i\!\in\!\{1,2,3,4\}$:    
\begin{subequations}
\begin{align}
\label{eq:LC1}
&F_1=-A_{1}\omega^{2}+\frac{\lambda^{(1)}N}{2}\mathcal{E}_{x}\\
\label{eq:LC2} 
&F_2=-A_{1}\omega\big[\alpha^{(1)} \!-\beta A_{1}^{2}\omega^{2}\big]+\frac{\lambda^{(1)}N}{2}\mathcal{E}_{y}\\
\label{eq:LC3}
&F_3=-A_{2}\omega\sin{\gamma}\big[\alpha^{(2)} \!-\beta A_{2}^{2}\omega^{2}\big]+A_{2}\omega^{2}\cos{\gamma} +\frac{\lambda^{(2)}N}{2}\mathcal{E}_{x}\\
\label{eq:LC4}
&F_4=A_{2}\omega\cos{\gamma}\big[\alpha^{(2)} \!-\beta A_{2}^{2}\omega^{2}\big]+A_{2}\omega^{2}\sin{\gamma} +\frac{\lambda^{(2)}N}{2}\mathcal{E}_{y}\\
\label{eq:LC5}
&\text{with} \nonumber \\
&\mathcal{E}_{x}=\int_{0}^{2\pi}\!\!\!\int_{0}^{R}\!\frac{A_{2}\cos{\gamma}+r\cos{\phi}-A_{1}}{d}\cdot\!\Big(\frac{C}{l}e^{-d/l}-e^{-d}\Big)\nonumber \\
&\;\;\;\;\;\;\;\;\;\;\;\;\;\;\;\;\;\;\cdot \frac{r dr d\phi}{\pi R^{2}}\\
\label{eq:LC6}
&\mathcal{E}_{y}=\int_{0}^{2\pi}\!\!\!\int_{0}^{R}\!\frac{A_{2}\sin{\gamma}+r\sin{\phi}}{d}\cdot\!\Big(\frac{C}{l}e^{-d/l}-e^{-d}\Big)\cdot \frac{r dr d\phi}{\pi R^{2}} \\ 
\label{eq:LC7}
&d=\sqrt{(A_{2}\cos{\gamma}+r\cos{\phi}-\!A_{1})^{2}+\!(A_{2}\sin{\gamma}+r\sin{\phi})^{2}}. 
\end{align}
\end{subequations}

\begin{figure}[t]
\center{\includegraphics[scale=0.212]{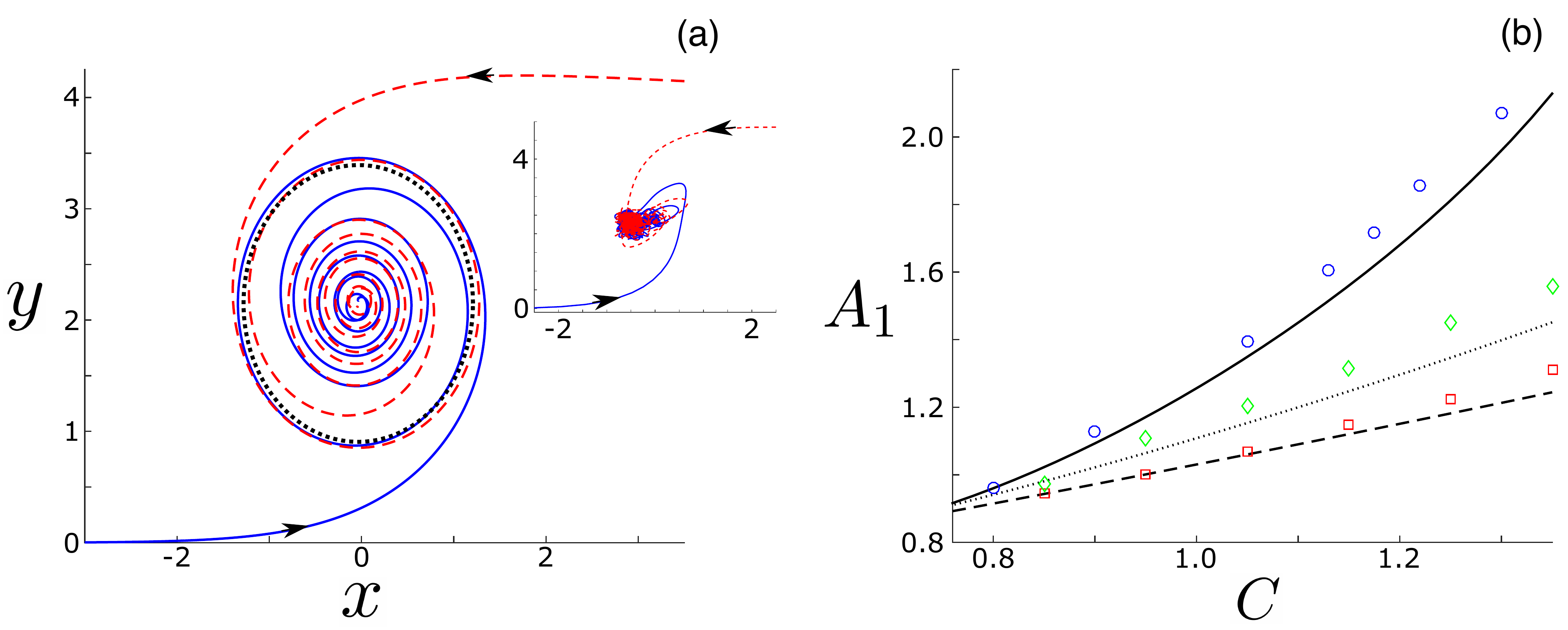}}
\caption{Collision dynamics resulting in milling. (a) Center-of-mass trajectories for two colliding swarms when $\lambda\!=\!\lambda_{min}$, shown with solid-blue and dashed-red lines. Arrows give the direction of motion. The dashed-black line indicates the bifurcating limit cycle in the uniform constant density approximation. Other swarm parameters are $\alpha\!=\!1$, $\beta\!=\!5$, $l\!=\!0.75$, $N\!=\!100$, and $C\!=\!1.0$. \corr{The inlet panel shows the corresponding trajectory for $\lambda\!=\!2 \lambda_{min}$}. (b) Maximum x-coordinate reached by the center of mass of the rightward moving (blue) flock when $\lambda\!=\!\lambda_{min}$. Simulation results are shown with blue circles for $l\!=\!0.75$, green diamonds for $l\!=\!0.6$, and red squares for $l\!=\!0.5$. Limit-cycle predictions from Eqs.(\ref{eq:LC1}-\ref{eq:LC4}) and Eq.(\ref{eq:SN}) are drawn with lines near each series. Other swarm parameters are $\alpha\!=\!1$, $\beta\!=\!5$, and $N\!=\!200$.}
\label{fig:Shedding3}
\end{figure}

Solutions to Eqs.(\ref{eq:LC1}-\ref{eq:LC4}) for ${\bm L}\equiv[A_1,A_2,\gamma,\omega]$ can be shown to exactly match limit cycles within the UCDA; more importantly, they agree with the transient oscillations for collisions in the full system, Eqs.(\ref{eq:SwarmModel}-\ref{eq:Morse}). For example, Fig.\ref{fig:Shedding3}(a) shows CM-trajectories in red and blue for two colliding swarms when $\lambda\!=\!\lambda_{min}$. We can see that the trajectories approach the UCDA limit-cycle, shown with a black-dashed line, before slowly decaying into the origin. \corr{The critical case can be contrasted to couplings above the critical value, e.g., $\lambda\!=\!2\lambda_{min}$ shown in the inlet panel, where the two colliding flocks rapidly decay into the MS}. Using this picture as a basis, the maximum rotation radius during collisions can be compared directly to limit-cycle radii predictions from Eqs.(\ref{eq:LC1}-\ref{eq:LC7}). In Fig.\ref{fig:Shedding3}(b) we plot such a comparison using the maximum horizontal distance reached by the CM of the rightward moving flock (as a proxy for the collision radius). UCDA predictions and simulations quantitatively agree fairly well over a broad range of parameter values. Qualitatively, as the repulsive-force constant $C$ increases, the two swarms oscillate at larger distances from each other upon collision, particularly for larger values of the repulsion scale, $l$. This increase in rotation distance, $A_{1}$, is accompanied by a decrease in rotation frequency, $\omega \sim A_{1}^{-1}$.

Next, we can consider stability. When control parameters are changed (one at a time), stable limit cycles satisfying Eqs.(\ref{eq:LC1}-\ref{eq:LC7}) disappear generically through {\it saddle-node bifurcations} (SNs).
As stated previously in Sec.\ref{sec:Collision}, a post-collision MS in the
full system Eqs.(\ref{eq:SwarmModel}-\ref{eq:Morse}) is not expected to form
unless stable limit-cycles exist, and hence, $\lambda_{min}$ can be
approximated by the SN value in the UCDA. We can find a general condition to
determine $\lambda_{min}$ at the SN through the following. 
Using the defined vector components $\bm F$ specified in
Eqs.(\ref{eq:LC1}-\ref{eq:LC4}), we compute the derivatives of $\bold{F}$
with respect to the limit-cycle parameters, $\bold{L}$. 
At the SN the Jacobian matrix $\underline{J}$, defined as $J_{mn}\!\equiv\!\partial F_{m}/\partial L_{n}$, has
\begin{align}
\det{\underline{J}(\bold{L};\lambda_{min})=0}. 
\label{eq:SN}
\end{align}
Combining Eq.(\ref{eq:SN}) with Eqs.(\ref{eq:LC1}-\ref{eq:LC4}) gives a total of $5$ root equations for the approximate critical coupling and associated limit-cycle. 

In practice, if we consider symmetric collisions or asymmetry in the
$\alpha$'s only (as we do in the remainder), the above results simplify. For
example, in the case of symmetric collisions the relevant branch of stable
limit cycles have $A_{1}\!=\!A_{2}$, $\gamma\!=\!\pi$, and
$\omega\!=\!\sqrt{\alpha/\beta}/A_{1}$. Moreover, the symmetric critical coupling predicts a scaling collapse \corr{(see App.\ref{sec:App2} for derivation and further details)}:
\begin{align}
\label{eq:Symmetric}
&\frac{2\alpha}{\lambda_{min}N\beta}\;=\;A_{1}^{2}\!\!\int_{0}^{2\pi}\!\!\!\int_{0}^{R}\!\frac{2rdrd\phi}{\pi R^{2}}\cdot\frac{\frac{C}{l}e^{-d/l}-e^{-d}}{d}\cdot\nonumber \\
&\Bigg[1-\frac{(2A_{1}-\!r\!\cos{\phi})^2}{d^{2}}-\frac{(2A_{1}-\!r\!\cos{\phi})^2}{d}\!\cdot\!\frac{\frac{C}{l^{2}}e^{-d/l}-e^{-d}}{\frac{C}{l}e^{-d/l}-e^{-d}}\Bigg]\!. 
\end{align}
where the right hand side is a function of the pairwise-interaction parameters only. \corr{The Eq.(\ref{eq:Symmetric}) is similar in structure to an escape velocity equation, e.g., from a fixed potential well
\begin{align}
\label{eq:escape}
v^{2}/2 -N\lambda_{min} V_{\text{eff}}(C,l)=0,
\end{align}
where $v$ is the speed of each flock, and $V_{\text{eff}}(C,l)$ quantifies the strength of the potential between agents (see App.\ref{sec:App2}). As a consequence, if the potential-forces and number of agents are held constant, flocks moving twice as fast require four times the coupling in order to be captured in a MS. Similarly, flocks with twice as many agents must fly $\sqrt{2}$-times faster in order to escape forming a MS.} 

\begin{figure}[h]
\center{\includegraphics[scale=0.194]{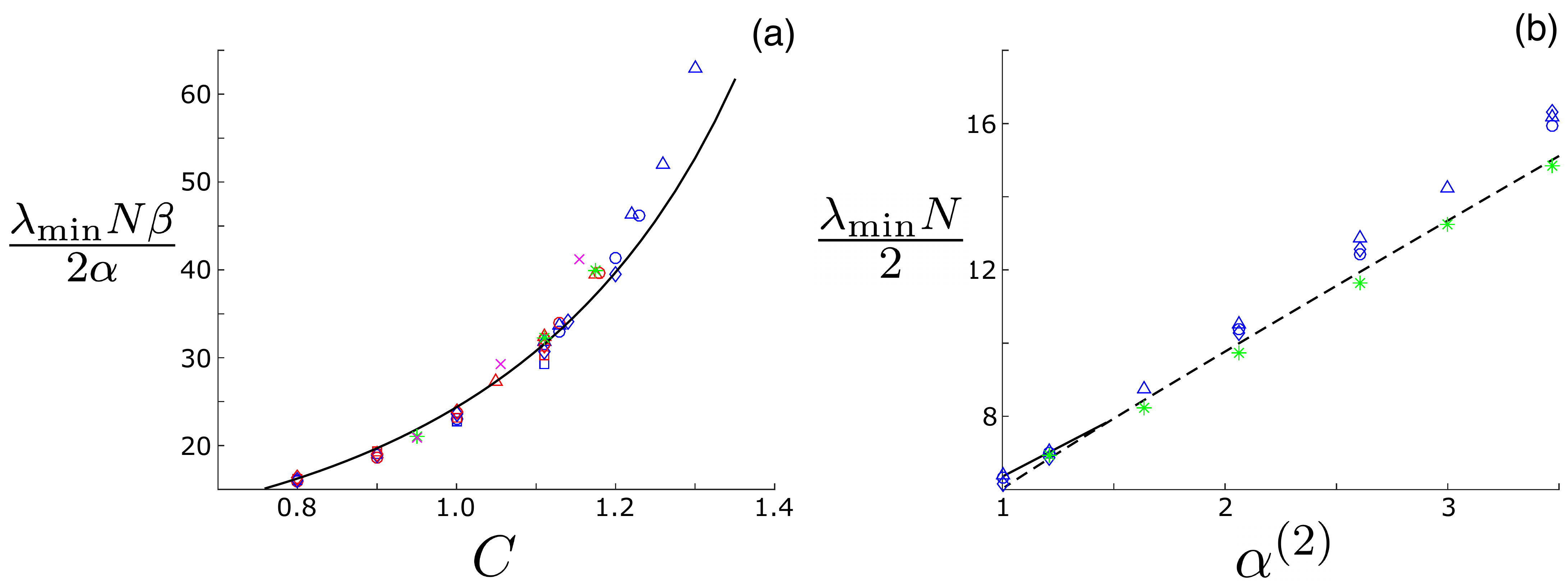}}
\caption{Critical coupling for forming milling states upon collision. (a)
  Symmetric parameter collisions for $\alpha\!=\!1$ (blue) and $\alpha\!=\!2$
  (red): $N\!=\!10$ (squares), $N\!=\!20$ (diamonds), $N\!=\!40$ (circles),
  and $N\!=\!100$ (triangles). Green stars denote $\alpha\!=\!1$ and magenta
  x's denote $\alpha\!=\!2$, when 40 agents collide with 60. (b) Asymmetric
  collisions for $C\!=\!10/9$ in which $\alpha^{(1)}\!=\!1$. Blue points
  indicate equal numbers in each flock: $N\!=\!20$ (diamonds), $N\!=\!40$
  (circles), and $N\!=\!100$ (triangles). Green stars denote collisions
  between 40 agents with $\alpha^{(1)}\!=\!1$ and $60$ agents with
  $\alpha^{(2)}$. Solid and dashed lines indicate theoretical predictions for (a)
  and (b), respectively from solving Eqs.(\ref{eq:LC1}-\ref{eq:LC4}) and Eq.(\ref{eq:SN}). Other swarm parameters are $\beta\!=\!5$ and
  $l\!=\!0.75$.}
\label{fig:Shedding4}
\end{figure}

Comparisons between $\lambda_{min}$ \corr{from simulations of the full discrete-particle system Eq.(\ref{eq:SwarmModel})}, and the above predictions from Eq.(\ref{eq:SN}) and Eqs.(\ref{eq:LC1}-\ref{eq:LC4}), are shown in Fig.\ref{fig:Shedding4}; $\lambda_{min}$ was measured by building final-scattering diagrams like Fig.\ref{fig:Shedding1}(a) for each parameter value. \corr{For these theory-simulation comparisons, note that the parameters $\alpha$, $N$, and $C$ are swept over a large range.} In the left subplot (a), we show results for collisions with symmetric parameters. As demonstrated with Eqs.(\ref{eq:Symmetric}-\ref{eq:escape}) our predicted scaling collapse holds. Qualitatively, the critical coupling increases monotonically with $C$, implying that the stronger the strength of repulsion, the larger the coupling needs to be in order for colliding swarms to form a MS. Also, note that our UCDA predictions are fairly robust to heterogeneities in the numbers in each flock, particularly for smaller values of $C/l-1$; predictions remain accurate for number asymmetries in the flocks as large as $20\%$.       

On the other hand, in Fig.\ref{fig:Shedding4}(b) we compare the measured $\lambda_{min}$ and predictions as a function of asymmetry in the self-propulsion force constant \corr{for different $N$'s}. The first swarm has $\alpha^{(1)}\!=\!1$, while $\alpha^{(2)}$ is varied. Contrary to the symmetric case the scaling collapse disappears, apart from $N$. Moreover, the branch of stable limit cycles with equal radii $A_{1}\!=\!A_{2}$ disappears in a cusp bifurcation (the solid-black line in in Fig.\ref{fig:Shedding4}(b) vanishes for $\alpha^{(2)}\!\gtrsim\!1.5$). Above the cusp point, the upper branch of SNs corresponds to stable limit cycles where $A_{1}\!<\!A_{2}$ and $\gamma\!=\!-\pi/2$, shown with a dashed-black line in the lower left corner of Fig.\ref{fig:Shedding4}(b). Interestingly, we can see that for larger values of $\alpha^{(2)}\!-\!\alpha^{(1)}$ the critical coupling is nearly linear in the difference, meaning that if one flock doubles its speed, then the coupling needed to form a MS is expected to quadruple -- again, a consequence of the flock speed equalling $\sqrt{\alpha/\beta}$. Finally, note that as in (a), predictions remain accurate for a significant range of differences in the numbers in each flock.

\section{\label{sec:DISCUSSION}DISCUSSION}
To summarize, in this work we studied the collision of two swarms with nonlinear interactions, and focused in particular on predicting when 
such swarms would combine to form a mill. Unlike the full final-scattering diagram, which depends on whether or not 
a particular set of initial conditions falls within the high-dimensional basin-of-attraction for milling -- a hard problem in general, we concentrated 
on predicting the minimum coupling needed to sustain a mill after the collision of two flocks. By noticing that colliding swarms, which eventually form a mill, 
initially rotate around a common center with an approximately constant density, we were able to transform the question of a critical coupling into 
determining the stability of limit-cycle states within a rigid-body approximation. \corr{This approach produced predictions that only depended on physical swarm parameters, 
and provided a lower-bound on the critical coupling for arbitrary impact parameters in nearly aligned colisions. For example, in the case of symmetric flocks with equal numbers and physical parameters, the scatter-mill transition 
point was similar to an escape-velocity condition in which the critical coupling scaled with the squared-speed of each flock, and inversely with the number of agents in each flock.} Our bifurcation analysis agreed well with many-agent simulations.

Though our analysis dealt directly with soft-core interacting swarms, the basic approach could
be extended to a broader range of models, as long as the forces between agents in Eq.(\ref{eq:SwarmModel}) have a finite range.
\corr{For instance, the results presented are similar for other choices of potential functions, which quantify the interaction between two agents, 
e.g., elastic interactions mediated through a network topology with an exponentially decaying coupling \cite{HindesShedding2021}. 
In terms of quantitative improvement in the calculation of the critical coupling, one straightforward approach would be to move beyond the 
uniform-density assumption, and replace the formulas in Secs.\ref{sec:UCDA}-\ref{sec:Oscillations} with an exact steady-state density for flocking states given an arbitrary choice of potential functions. 
As pointed out in Sec.\ref{sec:UCDA}, the UCDA works well, particularly for predicting the the sizes of flocks, but can both over and under predicts the 
density of agents near a flock's boundary (see Fig.\ref{fig:Shedding2}). Instead of assuming that upon collision agents 
maintain their relative configuration inside a flock, one can build an expansion in terms of the relative velocity of, e.g., the $i$th 
agent with respect to the flock's average, $\bold{v}_{i}\!=\! \left<\bold{v}\right>+\delta \bold{v}_{i}$, and keep the lowest order in $\delta\bold{v}_{i}(t)$.
Similar to the approach here, one can then try to analyze the stability of bound-state oscillations of two flocks that include the averaged dynamics and 
fluctuations around it, as in\cite{Mier_y_Teran_Romero_2014}. We also note that many other dynamical models of swarming, such as velocity-consensus models\cite{4200853}, produce 
flocking states that have a steady-state density of agents; we hypothesize that our analysis in terms of a critical coupling, slowly varying flocking densities, and 
bound-state oscillations may carry over to such models as well.} 

\corr{Beyond these generalizations, one can develop a similar approach for a broader range 
of swarm collision problems, since it is well known that systems like Eq.(\ref{eq:SwarmModel}) produce collective motion states other than flocking. 
For instance, one can develop an analogous dynamical system for analyzing the stability of collisions between different collective-motion states, such as flocks colliding with
mills. Typically, milling states also have a steady-state density of agents, even though individuals perform complex rotations around a stationary center\cite{Levine,DOrsagna,AlbiStability2014,Bernoff,Erdmann}. Here again, we believe our approach could prove useful.} 
  
\corr{In terms of applications, recent work in understanding the physics of swarm robotics and autonomy has begun to address how one swarm can detect, capture, redirect, 
or otherwise defend itself against another\cite{8444217,9029573,9303837}. Most current approaches, however, are primarily algorithmic, and lack 
basic physical and analytical insights. It is to this deficit that this work is in part addressed. In particular, our work fits nicely into the robotic swarm capture and redirect problem, since the critical coupling sets a general divide in parameterÊspace between scattering and milling swarms. For the collision of swarms of mobile robots, issues of noise, delay, and topology naturally arise, which are not included in this analysis. 
On small time scales we expect noise on Eq.(\ref{eq:SwarmModel}) to generate fluctuations in a flock's heading direction with  
a quasi-stationary probability distribution of agents in the co-moving frame, which has a larger boundary than predicted 
by Eq.(\ref{eq:R}). In the presence of noise, it is this distribution which would replace the density in    
the formulas in Secs.\ref{sec:UCDA}-\ref{sec:Oscillations}. We point out that, since noise will inevitably cause drift in real 
collisions, our lower-bound stability result becomes perhaps more useful, not less, since the exact collision distance $\Delta y$  
cannot be controlled deterministically. On longer time-scales the two flocks will lose agents, one by one, in a process that is similar to 
noise-induced escape; the escape of individual agents is much more likely in flocks with finite-range interactions than switching, since the 
latter requires a collective-fluctuation of many agents simultaneously\cite{6580546}. In Fig.\ref{fig:Shedding4} we show that our critical-coupling results 
are fairly robust to modest variations in the number of agents within each flock, and hence we do not expect the noise-induced loss of agents to significantly change our results.
On the other hand, the effect of delay in swarming dynamics can produce stable oscillations in the center of mass of a swarm\cite{hindes2020stability}. In fact, simulation results similar to those shown in Fig.\ref{fig:Shedding1} have suggested that time-delayed interactions tend to expand the red-region, effectively reducing the critical coupling at which bound-state oscillations become stable\cite{kolon2018dynamics}. Finally, communication topology significantly affects robotic swarming dynamics, for instance, generating new kinds of hybrid motion-states and transitions.
However, mean-field techniques that properly account for topology, yet are similar to those deployed here, have been shown to provide quantitatively accurate insights on the role of topology in determining swarm dynamics\cite{hindes2016hybrid,HindesShedding2021}. Future robotics experiments, similar to \cite{szwaykowska2016collective,edwards2020delay,4209425}, will be used to further test and expand our analysis in these and other scenarios.}



\section{\label{sec:Acknowledgments}ACKNOWLEDGMENTS}
JH and IBS were supported by the U.S. Naval Research Laboratory funding
(N0001419WX00055), the Office of Naval Research (N0001419WX01166) and
(N0001419WX01322), and the  Naval Innovative Science and Engineering. 
TE was supported through the U.S Naval Research Laboratory
Karles Fellowship.\\


\vspace*{-0.8cm} 
\corr{\section{\label{sec:App}APPENDIX}}
\corr{\subsection{\label{sec:App1}Two-agent milling}}
\corr{In the simple two-agent case, we can calculate when the red milling region in Fig.\ref{fig:Shedding1}(b) first emerges without resorting to approximations. When the two agents have equal parameters, milling consists of a circular-orbit limit cycle with $\bold{r}^{(1)}(t)\!=\!a\cos(\omega t)\hat{\bold{x}}+a\sin(\omega t)\hat{\bold{y}}$ and $\bold{r}^{(2)}(t)\!=\!-a\cos(\omega t)\hat{\bold{x}}-a\sin(\omega t)\hat{\bold{y}}$. Substituting this ansatz into Eqs.(\ref{eq:SwarmModel}-\ref{eq:Morse}) gives the following relation for the limit-cycle amplitude $a$:} 
\corr{\begin{align}
\label{eq:TAa}
0=\frac{\alpha}{\beta\lambda a} +\frac{C}{l}e^{-2a/l} -e^{-2a}. 
\end{align}}

\corr{The limit-cycle disappears at a SN bifurcation corresponding to a critical amplitude, $a^{*}$. Applying the zero-determinant condition, Eq.(\ref{eq:SN}), results in} 
\corr{\begin{align}
\label{eq:TASN}
0=-2a^{*} +\frac{\frac{C}{l}e^{-2a^{*}\!/l} -e^{-2a^{*}}}{\frac{C}{l^{2}}e^{-2a^{*}\!/l} -e^{-2a^{*}}}.
\end{align}} 
\noindent\corr{Note that $a^{*}$ only depends on $C$ and $l$, $a^{*}(C,l)$. Finally, combining Eqs.(\ref{eq:TAa}-\ref{eq:TASN}), gives the critical coupling} 
\corr{\begin{align}
\label{eq:TASNC}
\lambda_{min}=\frac{\alpha}{\beta a^{*}\big[\frac{C}{l}e^{-2a^{*}\!/l} -e^{-2a^{*}}\big] }.
\end{align}}
\corr{\noindent For reference in Fig.\ref{fig:Shedding1}(b) $\lambda_{min}=5.476$, which agrees with the Eq.(\ref{eq:TASNC}) solution $\lambda_{min}=5.473$, within the resolution of the simulations.} 

\corr{Note that $\lambda_{min}$ gives a lower bound for the scatter-mill transitions implied in Fig.\ref{fig:Shedding1}(b) for every $\Delta y$, since scattering cannot produce milling if no milling solution is locally stable.}\\ 
 
\corr{\subsection{\label{sec:App2}Symmetric collision scaling}}
\corr{In this section we further discuss the derivation and scaling of Eq.(\ref{eq:Symmetric}). As described in Sec.\ref{sec:Oscillations}, a stable limit-cycle exists within the UCDA for two flocks with equal parameters, $\alpha, \beta, C, l,\lambda, \text{and}\; N$, as long $\lambda\!>\!\lambda_{min}$. In this symmetric case, the limit cycle has amplitude(s) $A_{1}\!=\!A_{2}$ and relative phase $\gamma\!=\!\pi$. Applying Eqs.(\ref{eq:LC1}-\ref{eq:LC4}) results in the following two relations:
\begin{align}
\label{eq:MoreDetail1}
&A_{1}\omega^{2}=\frac{\lambda N}{2}\mathcal{E}_{x}(A_{1},C,l),\\
\label{eq:MoreDetail2}
&A_{1}^{2}\omega^{2}=\alpha/\beta. 
\end{align}
Note that the function $\mathcal{E}_{x}(A_{1},C,l)$ does not depend on the flock boundary, $R$, explicitly since $R(C,l)$ is determined by Eq.(\ref{eq:R}). Eliminating $\omega$ from Eqs.(\ref{eq:MoreDetail1}-\ref{eq:MoreDetail2}) gives
\begin{align}
\label{eq:MoreDetail3}
\frac{2v^{2}}{\lambda N}=A_{1}\mathcal{E}_{x}(A_{1},C,l),
\end{align}
where we have used the velocity for the flocking state $v^{2}\!=\!\alpha/\beta$.} 

\corr{Next, the general saddle-node condition, Eq.(\ref{eq:SN}), reduces to setting a single derivative of Eq.(\ref{eq:MoreDetail3}) equal to zero,  
\begin{align}
\label{eq:MoreDetail4}
0=\frac{\partial}{\partial_{A_{1}}}\;A_{1}\;\mathcal{E}_{x}(A_{1},C,l).
\end{align}
Solving Eq.(\ref{eq:MoreDetail4}) for $A_{1}$ gives the critical amplitude $A_{1}^{*}$, which only depends on $C$ and $l$, or $A_{1}^{*}(C,l)$. Finally, combining Eq.(\ref{eq:MoreDetail3}), evaluated at the critical point, with Eq.(\ref{eq:MoreDetail4}) results in the following expression equivalent to Eqs.(\ref{eq:Symmetric}-\ref{eq:escape}):  
\begin{align}
\label{eq:MoreDetail5}
\frac{2v^{2}}{\lambda_{min} N}=A_{1}^{*}(C,l)\mathcal{E}_{x}(A_{1}^{*}(C,l),C,l).
\end{align}}

\corr{Equation (\ref{eq:MoreDetail5}) implies that $v^{2}\big/\lambda_{min} N$ is a function only of the interaction-force constants, $C$ and $l$, and hence $\lambda_{min}\sim v^{2}/N$.}  


\begin{thebibliography}{65}
\bibitem [{Not()}]{NoteColl2}%
  \BibitemOpen
  \href@noop {} {}\bibinfo {note} {We note that a similar approximation can be
  made in lieu of Eqs.(5), in terms of the center-of-mass of each flock,
  $\bold{R}^{(k)}(t)\equiv\sum_{i\in(k)}\bold{{r}}_{i}/N_{k}$. A consequence of
  the UCDA approximation is that
  $\dot{\bold{R}}^{(k)}\!\approx\!\dot{\bold{r}}_{i}\;\forall i\in(k)$ (the
  flocks keep their relative configurations), and therefore the equation for
  $\ddot{\bold{R}}^{(k)}$ only differs from Eqs.(5) in the interaction sums. In
  $\ddot{\bold{R}}^{(k)}$, however, four integrals appear-- instead of
  two--since one must sum over pairs of agents in both flocks. Because of this
  additional complexity, and the fact that the dynamics for both approaches
  have only slight numerical differences, we restrict our analysis to
  Eqs.(5)}\BibitemShut {NoStop}%
\end{thebibliography}
%

\end{document}